# A signature of power law network dynamics

Ashish Bhan* and Animesh Ray*

Center for Network Studies

Keck Graduate Institute

535 Watson Drive

Claremont, CA 91711

*Authors for correspondence (emails: abhan111@gmail.com; aray@kgi.edu)

Can one hear the 'sound' of a growing network? We address the problem of recognizing the topology of evolving biological or social networks. Starting from percolation theory, we analytically prove a linear inverse relationship between two simple graph parameters—the logarithm of the average cluster size and logarithm of the ratio of the edges of the graph to the theoretically maximum number of edges for that graph—that holds for all growing power law graphs. The result establishes a novel property of evolving power-law networks in the asymptotic limit of network size. Numerical simulations as well as fitting to real-world citation co-authorship networks demonstrate that the result holds for networks of finite sizes, and provides a convenient measure of the extent to which an evolving family of networks belongs to the same power-law class.

Mark Kac posed the question "Can one hear the shape of a drum?" (1), asking whether one could infer the geometric shape of a drum from its vibrational modes. Here we explore a similar question about the evolution of complex networks. The structure and properties of complex networks (2-4), such as the Internet (5, 6), citation networks (7), biological interaction network such as metabolic (8), protein-interaction networks (9), and disease-gene network (10) have been investigated intensively over the past decade. Many of these real world networks belong to the power-law family where the number of nodes with $k$ neighbors, $N(k)$, scales as a power law of the form $N(k) \sim k^{-\beta}$ where $\beta$ is the *scaling coefficient (11)*. Several mechanisms by which power law networks arise through evolutionary mechanisms were proposed by considering the structure of the real-world network as a target and by constructing models of network growth that produce such a network given an arbitrary seed graph as the initial condition (12-15). An important problem that arises in evaluating models of network evolution and in testing these against real data is the need to definitively characterize experimentally determined networks as belonging to the power-law family (16). Here we consider a related problem—given a time-series of network structure data, can one deduce whether the collection of networks belongs to the same power-law class?

In percolation theory (17), the vertices of a lattice are "occupied" or connected with a certain probability $p$, and the emergence of connected clusters are studied as a function of $p$. Of interest is the existence of a path that would allow a liquid, as it were, to percolate from the upper surface of the lattice to its lower surface. Many of the canonical results in percolation theory are in the form of power laws $W \propto (p - p_c)^k$ where $p_c$ is a constant value denoting the critical probability at which a phase transition occurs, and $W$ is a parameter, such as the average cluster size being studied. There is no equivalent theoretical framework for evolving real-world networks that belong to a power-law family.

Consider a graph with $N$ nodes and $E$ edges. If $N(k)$ is the number of nodes of degree $k$, the associated probability distribution of nodes is given by $P(k) = \frac{N(k)}{N}$, where $K$ is the maximum degree. The average degree of the graph $\langle k \rangle = \sum_{k=1}^{K} k \frac{N(k)}{N} = \sum_{k=1}^{K} k P(k) = \frac{2E}{N}$. The fraction of edges, $F$, relative to the number of all possible edges in the complete graph on $N$ nodes, is $\frac{E}{\binom{N}{2}} = \frac{2E}{N(N-1)} = \frac{\langle k \rangle}{N-1}$. We define $W_k$, the likelihood that an edge chosen at random is connected to a node of degree $k$, as $\frac{kN(k)}{\sum_{k=1}^{K} kN(k)} = \frac{kN(k)}{\langle k \rangle N} = \frac{k}{\langle k \rangle} P(k)$.

Using a term from percolation theory we define the **average cluster size** $W \equiv \sum_{k=1}^{K} W_k N(k)$. The generalized moments $\lambda_{n,m}$ are given by the expression $\sum_{k=1}^{K} k^n P(k)^m$. Note that $\lambda_{1,1} = \langle k \rangle$. For networks where $K$ is large, it can be proved that $W \cdot F = \frac{\beta - 1}{2}$ (see analytical derivation in **Methods**).

The analysis above demonstrates that for power-law networks that belong to the same class $W$ scales as $1/F$ in the limit of large $N$. We now demonstrate by means of simulations that this scaling law holds for power-law networks of finite size (Fig 1A) but not to Erdös-Renyi random networks (Fig 1B). The scaling of $W$ as a function of $F$ can be used to discriminate sensitively between types of networks produced by previously described network growth models by full or partial node duplication (15). The graphs in Figs 2A and 2B correspond to networks constructed according to a partial duplication model, which produces power-law networks for values of $p$, $p < \frac{\sqrt{5}-1}{2}$ (7, 14, 15, 18). As $p$ approaches 1, the degree distribution begins to deviate from a power-law distribution. This is shown by the graphs in Figs 2C and 2D, which show the progressive deviation from the power-law as reflected both in the graphs of $N(k)$ vs. $k$ and $W$ vs. $F$ on the log-log scale.

We have tested for the scaling law described above in data from the Citeseer database (19, 20). For each year between 1991 and 1999, we extracted the list of papers published and the number of citations to each paper and plotted the degree distribution for citation network for each year (Fig 3A). The degree distributions appear to fit poorly a power-law. In Fig 3B we plot $W$ as a function of $F$ for these 9 networks, and on the log-log scale $W$ is approximately a linear function of $F$ with a slope of -0.86 (close to -1 as predicted by our analysis).

The result described above for the average cluster size $W$ varying as an inverse function of $F$ does not discriminate among different types of power-law networks—it identifies growing networks in this broad class. Here we only study the behavior of the moment $\lambda_{1,2}$ as a function of the fraction of possible edges $F$. The moments that we define in equation 2 are analogous to the moments of a probability distribution, and may serve as a rigorous method for characterization of the structure of complex networks and their modes of evolution. This simple scaling law should aid diverse purposes where network based simulation is important, such as finding patterns in social networks and for epidemiological modeling to optimize immunization strategies. Moreover, modularization of large and complex networks into component subgraphs, each with a power-law topology, might also be possible.

**Acknowledgements**

This work is supported by EAGR-0941078 (NSF), FIBR-0527023 (NSF), and 1R01GM084881-01 (NIH) grants to AR.

**Figure Legends:**

**Figure 1**. Characteristic parameters that describe the growth features of power law networks. Beginning with a fully connected graph on 3 nodes and growing it according to the preferential attachment scheme of Barabsi and Albert (11) with $m_0 = 2$ and for *t = 100, 200, 300, …2000*, twenty networks of increasing sizes were produced. The degree distributions of these networks approach a power-law in the asymptotic limit (11). For each of these twenty networks we computed *W* and *F* (panel A). This is a considerable reduction of complexity: where we once had twenty different degree distributions, each of varying goodness of fit to the ideal power-law distribution, we now have twenty points whose colinearity in a log-log plot can be easily checked. In panel B, we simulated 20 Erdos-Renyi random networks with 100 nodes where $F$ was varied in steps of size 0.01 from 0.4 to 0.6. For each of these random networks we measured the average cluster size *W* and plotted as a function of *F*.

**Figure 2** Characteristic parameters describe network growth models. Networks were produced using the partial node and edge duplication model (15) for *p = 0.1, 0.2, 0.3,* and *0.4* (panels A, C), and for *p = 0.7, 0.8, 0.9,* and *0.95*, respectively (B, D). For *p* < 0.5, the model is expected to produce power law network but for *p* > 0.5 the networks progressively deviate from the power law.

**Figure 3** To test whether the derived property holds for real networks, citation data of papers published during 1991-1999 were analyzed. For each year, we built a citation network and plotted the degree distribution, *N(k)*, as a function of *k* in panel (A). The degree distributions appear to be noisy power-laws. For each of the 9 networks in panel (A), we computed *W* and *F*, and plotted these on a log-log scale in panel (B), showing that the growth property indeed belongs to the power law family. The noisy inverse relationship we observe characterizes the networks as those growing in the same power-law family, but in a noisy manner.

---

**Methods**

Analytical Derivation: For networks where the maximum degree, $K$ is large, we can approximate the sum in the definition of average cluster size *W* by an integral. Then, for $\beta \neq 1$, $\int_1^K P(k)dk = C\int_1^K k^{-\beta}dk = C\left(\frac{1-K^{-(\beta-1)}}{(\beta-1)}\right)$.

The normalization constant, $C$, depends on $\beta$ and $K$ and is given by $C = \frac{(\beta-1)}{1-K^{-(\beta-1)}}$. Hence, for large $K$ and $\beta > 1$, the limiting value of $C$ is $\tilde{C} = \beta - 1$. Substituting in the expression for generalized moment (see text), we obtain $\lambda_{n,m} = \sum_{k=1}^{K} k^n P(k)^m \approx C^m \int_1^K k^{n-m\beta} dk = C^m \frac{K^{n-\beta m+1}-1}{n-\beta m+1}, n-\beta m \neq -1$. Using the normalized value of $C$ we can rewrite the above as $\lambda_{n,m} \approx C^m \left(\frac{K^{n-m\beta+1}-1}{n-m\beta+1}\right) = \left(\frac{(\beta-1)^m}{1-K^{-(\beta-1)}}\right)\left(\frac{K^{n-m\beta+1}-1}{n-m\beta+1}\right)$, which simplifies in the limit of large $K$ to $\Lambda_{n,m} \equiv \lim_{K\to\infty} \lambda_{n,m} = \frac{(\beta-1)^m}{(m\beta-n-1)}$.

The relationship between the quantities *W* and *F* is obtained from the definition of $F = \frac{\langle k \rangle}{N-1} = \frac{\lambda_{1,1}}{N-1}$ and

$$W \equiv \sum_{k=1}^{K} W_k N(k) = \frac{\sum_{k=1}^{K} k N(k)^2}{\sum_{k=1}^{K} k N(k)} = \frac{\sum_{k=1}^{K} k P(k)^2}{\sum_{k=1}^{K} k P(k)} N = \frac{\lambda_{1,2}}{\lambda_{1,1}} N.$$ In the case of large networks or in the limit as $K \to \infty$ we obtain, $W.F = \frac{N}{N-1} \Lambda_{1,2}$ or, $W.F = \frac{\beta-1}{2}$.

**References**

1. Kac M (1966) Can one hear the shape of a drum? *American Mathematical Monthly* 73(4):1-23.
2. Albert R, Barabasi, A. (2002) Statistical mechanics of complex networks. *Rev Mod Phys* 74:47-97.
3. Brown KS*, et al.* (2004) The statistical mechanics of complex signaling networks: nerve growth factor signaling. *Physical biology* 1(3-4):184-195.
4. Strogatz SH (2001) Exploring complex networks. *Nature* 410(6825):268-276.
5. Albert R, Jeong H, & Barabasi AL (2000) Error and attack tolerance of complex networks. *Nature* 406(6794):378-382.
6. Faloutsos M, Faloutsos P, & Faloutsos C (1999) On power-law relationships of the Internet topology. *Proceedings of the conference on Applications, technologies, architectures, and protocols for computer communication*, pp 251-262.
7. Girvan M & Newman ME (2002) Community structure in social and biological networks. *Proc Natl Acad Sci U S A* 99(12):7821-7826.
8. Jeong H, Tombor B, Albert R, Oltvai ZN, & Barabasi AL (2000) The large-scale organization of metabolic networks. *Nature* 407(6804):651-654.
9. Han JD*, et al.* (2004) Evidence for dynamically organized modularity in the yeast protein-protein interaction network. *Nature* 430(6995):88-93.
10. Goh KI*, et al.* (2007) The human disease network. *Proc Natl Acad Sci U S A* 104(21):8685-8690.
11. Barabasi AL & Albert R (1999) Emergence of scaling in random networks. *Science* 286(5439):509-512.
12. Foster DV, Kauffman, S.A., Socolar, J.E.S. (2006) Network growth models and genetic regulatory networks. *Phys Rev E* 73(3):031912-031920.
13. Albert R, Barabasi, A. (2000) Topology of evolving networks: local events and universality. *Phys Rev Lett* 85(24):5234-5237.
14. Chung F, Lu L, Dewey TG, & Galas DJ (2003) Duplication models for biological networks. *J Comput Biol* 10(5):677-687.
15. Bhan A, Galas DJ, & Dewey TG (2002) A duplication growth model of gene expression networks. *Bioinformatics* 18(11):1486-1493.
16. Clauset A, Shalizi, C.R., Newman, M.E.J. (2009) *SIAM Review* 51:661.
17. Stauffer D, Aharony, A. (1994) *Introduction to percolation theory* (CRC Press).
18. Kim J, Krapivsky PL, Kahng B, & Redner S (2002) Infinite-order percolation and giant fluctuations in a protein interaction network. *Phys. Rev. E* 66(055101(R)).
19. An Y, Janssen, J., Milios, E.E. (2004) Characterizing and mining the citation graph of the computer science literature. *Knowledge and Information Systems* 6(6):664-678.

20. An Y, Janssen J, & Milios EE (2004) Characterizing and mining the citation graph of the computer science literature. *Knowledge and Information Systems* 6(6):664-678.

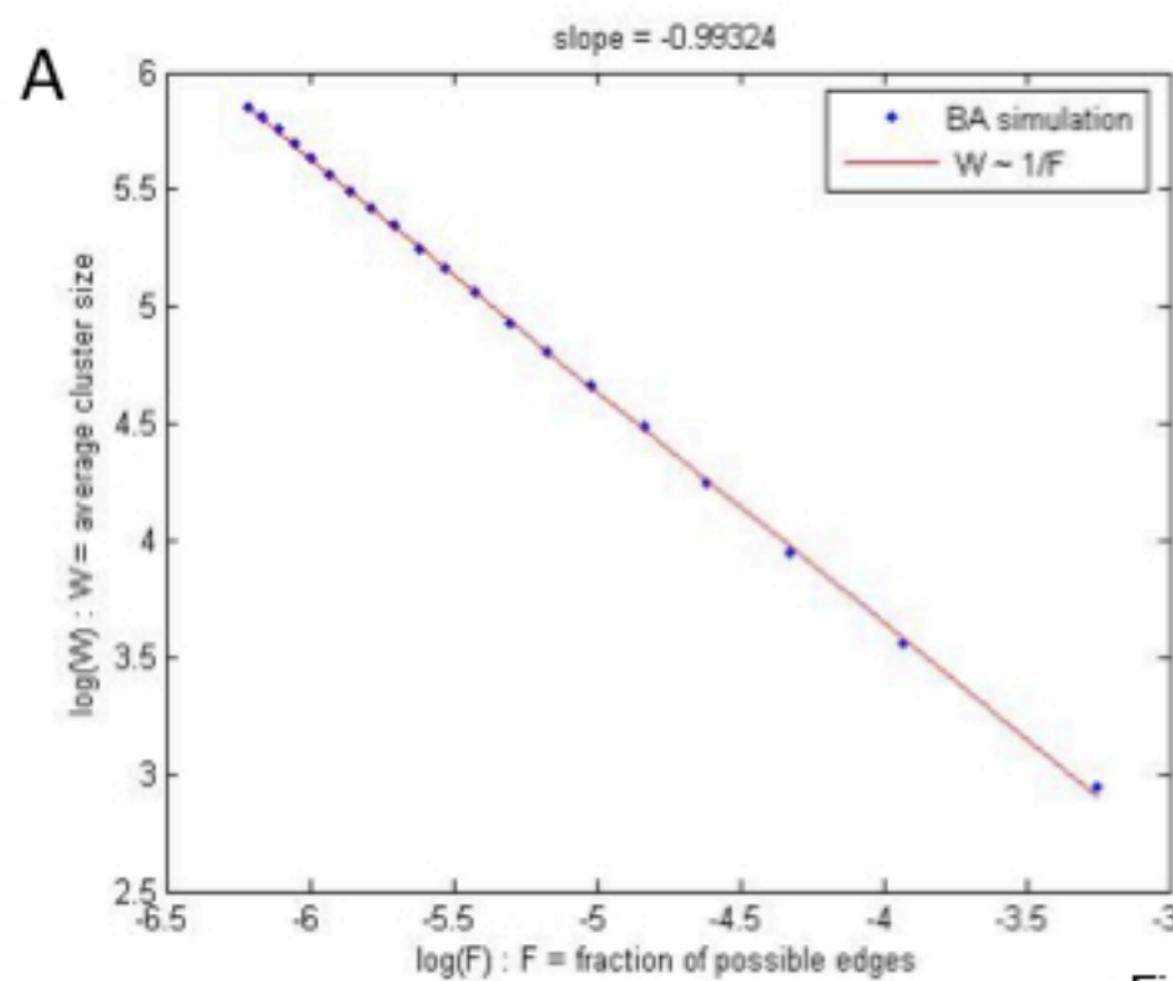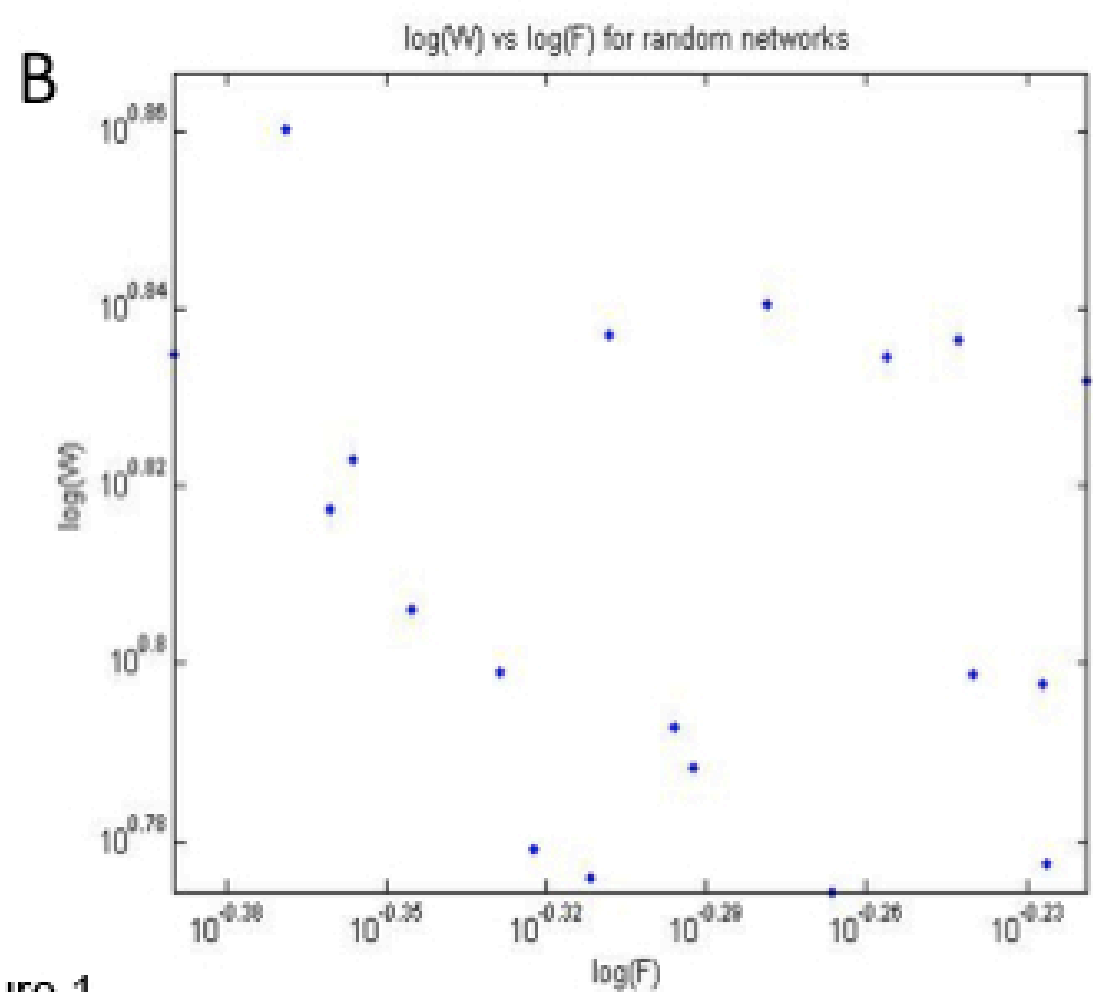

Figure 1

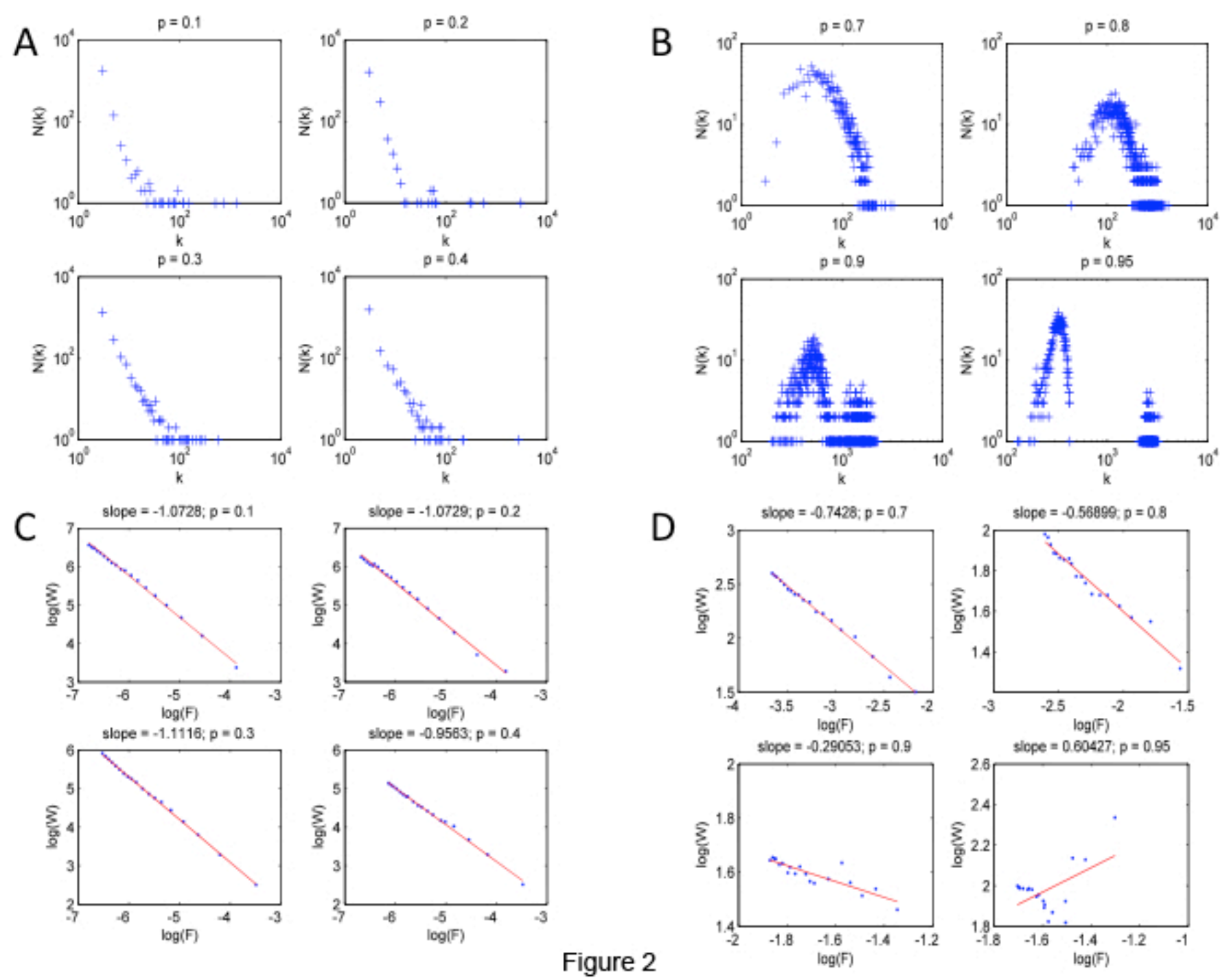

Figure 2

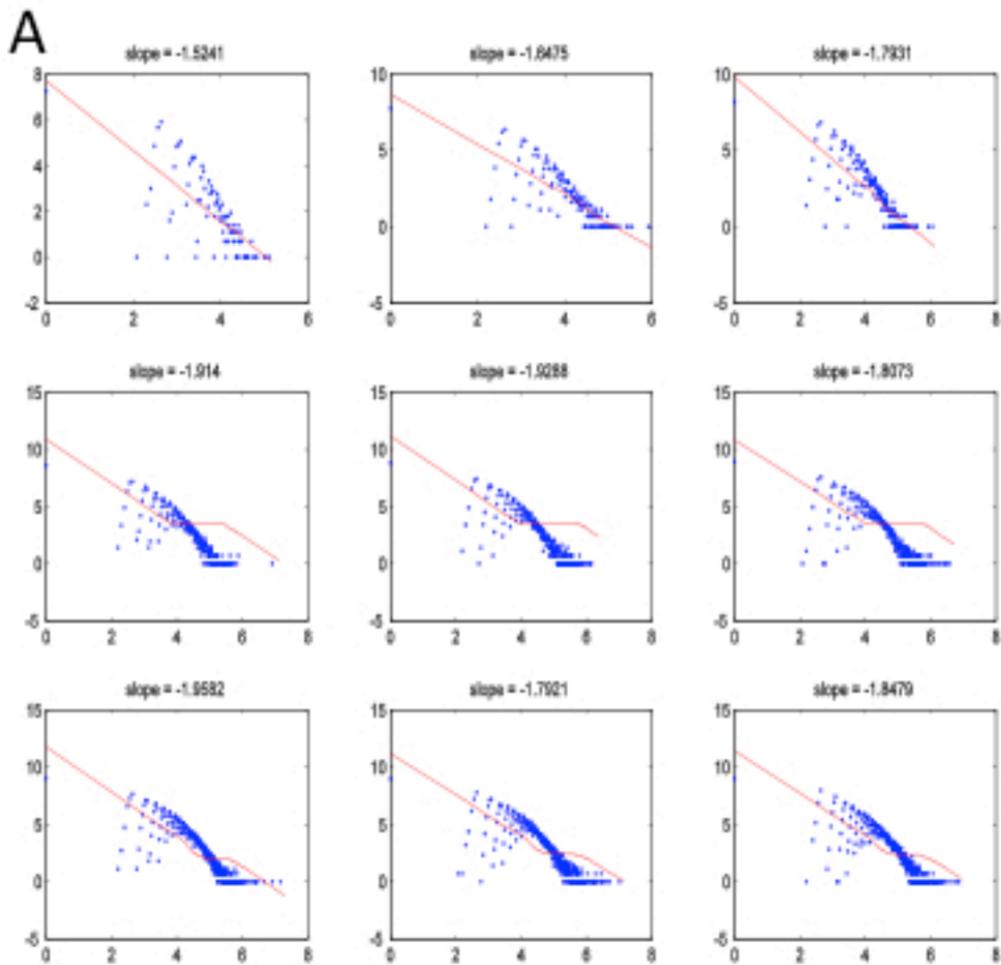
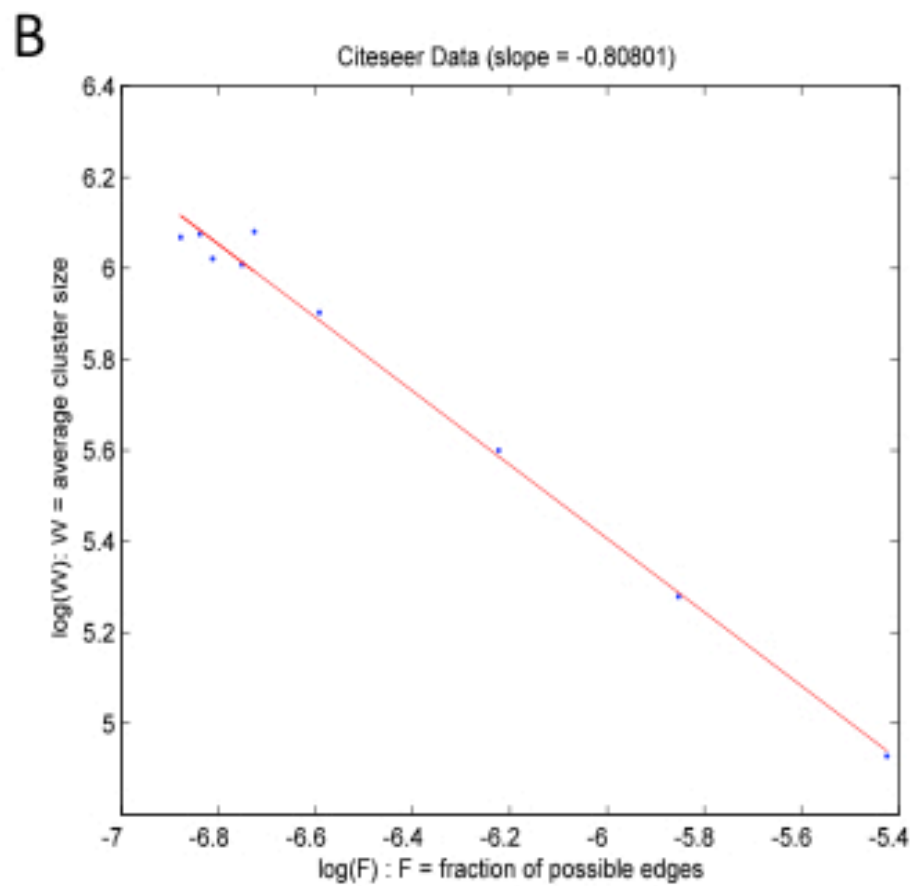

Figure 3